\documentclass[3p,times,twocolumn]{elsarticle}

\usepackage{amssymb}

\usepackage[figuresright]{rotating}

\begin{document}

\begin{frontmatter}




\title{Combined constraints on CP-violation in the Standard Model and beyond}


\author{S. Descotes-Genon$^{a,b}$ (on behalf of the CKMfitter group)}

\address{$^a$ Univ. Paris-Sud, Laboratoire de Physique 
Th\'eorique, UMR 8627,
91405 Orsay Cedex, France\\
$^b$ CNRS, 91405 Orsay, France}

\begin{abstract}
I review the status of CP violation in the Standard Model from the combination of flavour constraints within the CKMfitter frequentist approach and I describe studies of New Physics restricted to the $\Delta F=2$ sector to explain recent results on neutral-meson mixing. All results have been obtained using data available for the Winter 2012 conferences.
\end{abstract}




\end{frontmatter}





\section{Standard Model global fit}

In the Standard Model (SM), the weak charged-current transitions
mix different quark generations, which is encoded in the
unitary Cabibbo-Kobayashi-Maskawa (CKM) matrix.
In the case of three generations of quarks,
the physical content of this matrix reduces to four real parameters,
among which one phase, the only source of CP violation in the quark sector. These four real parameters are defined in an phase-convention independent way:
\begin{eqnarray}
\lambda^2&=&\frac{|V_{us}|^2}{|V_{ud}|^2+|V_{us}|^2}
\qquad
A^2\lambda^4 =\frac{|V_{cb}|^2}{|V_{ud}|^2+|V_{us}|^2}\nonumber\\
\!\!\!\!\!\!\bar\rho+i\bar\eta&=&-\frac{V_{ud}V_{ub}^*}{V_{cd}V_{cb}^*}\,.
\end{eqnarray}
As these parameters have been determined more and more precisely over the last decade, the attention has shifted from the pure metrology of the Standard Model to the investigation of deviations in flavour physics, tell-tale signs of New Physics (NP). The CKMfitter group follows this programme within a frequentist approach, using the Rfit model to describe systematic uncertainties~\cite{ckmfitter}. A table of the inputs summarising the SM global fit is presented in Table~\ref{tab:inputs}. Low-energy strong interactions constitute 
a central issue in flavour physics, which explains the need for accurate inputs for hadronic quantities such as decay constants, form factors, bag parameters\ldots We mostly rely
on Lattice QCD simulations, with a specific averaging procedure (Educated Rfit) to combine the results from different collaborations. 
A similar approach is followed in order to combine the inclusive and exclusive determinations of $|V_{ub}|$ and $|V_{cb}|$.

\begin{table*}[t]
\begin{center}
\begin{tabular}{|c|l|l|}
\hline
Quantity & Experimental information & Theoretical input \\
\hline
$|V_{ud}|$   & superallowed $\beta$ decays & Ref.~\cite{Hardy:2008gy}\\
$|V_{us}|$   & WA $K_{\ell 3}, K_{\ell 2}, \pi_{\ell 2}, \tau\to K\nu_\tau, \tau\to \pi\nu_\tau$
  & $f_+(0)=0.963\pm 0.003\pm 0.005$\\
$\epsilon_K$ & WA & $\hat{B}_K=0.733\pm 0.003\pm 0.036$ \\
$|V_{ub}|$   & $b\to u\ell\nu$, $B\to \pi\ell\nu$ & 
$|V_{ub}|\cdot 10^3= 3.92 \pm 0.09 \pm  0.45$ 
 \\
$|V_{cb}|$   & $b\to c\ell\nu$, $B\to D^{(*)}\ell\nu$ & 
$|V_{cb}|\cdot 10^3= 40.89\pm 0.38 \pm 0.59$
 \\
$\Delta m_d$ & WA $B_d$-$\bar{B}_d$ mixing & $B_{B_s}/B_{B_d}=1.024 \pm 0.013\pm 0.015$ \\
$\Delta m_s$ & WA $B_s$-$\bar{B}_s$ mixing & $B_{B_s}=1.291 \pm 0.025 \pm 0.035$ \\
$\beta$  &  WA $J/\psi K^{(*)}$ & \\
$\alpha$ &  WA $\pi\pi,\rho\pi,\rho\rho$ & isospin\\
$\gamma$ & WA $B \to D^{(*)} K^{(*)}$   & GLW/ADS/GGSZ\\
$BR(B \to \tau \nu)$ & $(1.73\pm 0.35)\cdot 10^{-4}$ & $f_{B_s}/f_{B_d}=1.218\pm 0.008 \pm 0.033$\\
&& $f_{B_s}=229 \pm 2 \pm 6$ MeV   \\
\hline
\end{tabular}
\caption{Summary of the inputs for the SM global fit. WA stands for "World average" from PDG or HFAG. For further details, see Ref.~\cite{ckmfitter} and references therein.\label{tab:inputs}}
\end{center}
\end{table*}

An input which has varied recently is the angle $\gamma$, whose
 measurement is based on the interference between the colour-allowed 
$B^- \to D^0 K^-$ and colour-suppressed $B^- \to \bar{D}^0 K^-$
decays (as well as modes with similar flavour structure). The accuracy of the method is driven by the size of 
the ratio $r_B = |A_{\rm suppressed}|/|A_{\rm allowed}|$ typically of order 0.1-0.2. The different
methods try to improve on this ratio by different choices of
$D$ decay channels: GLW ($D$ into CP eigenstates), ADS ($D^{(*)}$ into doubly Cabibbo-suppressed states), GGSZ ($D^{(*)}$ into 3-body state and Dalitz analysis). Each method allows for a simultaneous determination of $\gamma$ and the relevant hadronic quantities, in particular the ratio of amplitudes $r_B$ and the relative phase between the suppressed and allowed amplitudes $\delta_B$.
Recent measurements for the ADS method from CDF~\cite{Aaltonen:2011uu} and Belle~\cite{BELLE-CONF-1112} in Summer 2011 and
from LHCb~\cite{Aaij:2012kz} in Winter 2012
together with 
a more sophisticated statistical treatment of the nuisance parameters (constrained supremum) have led to
a significant improvement for the input of the global fit corresponding to the combination of the three methods.

\begin{equation}
\!\!\!\!\!\!\!\!\begin{array}{cccc}
\gamma(^\circ) & {\rm Summer\ 10} & {\rm Summer\ 11} & {\rm Winter\ 12} \\ 
{\rm fit\ input} & 71^{+21}_{-25} & 68^{+10}_{-11} & 66^{+12}_{-12}\\
{\rm fit\ output} & 67.2^{+3.9}_{-3.9} & 67.3^{+4.2}_{-3.5} & 67.1^{+4.3}_{-4.3}
\end{array}
\end{equation}

\begin{figure*}[t]
\begin{center}
\includegraphics[height=.33\textheight]{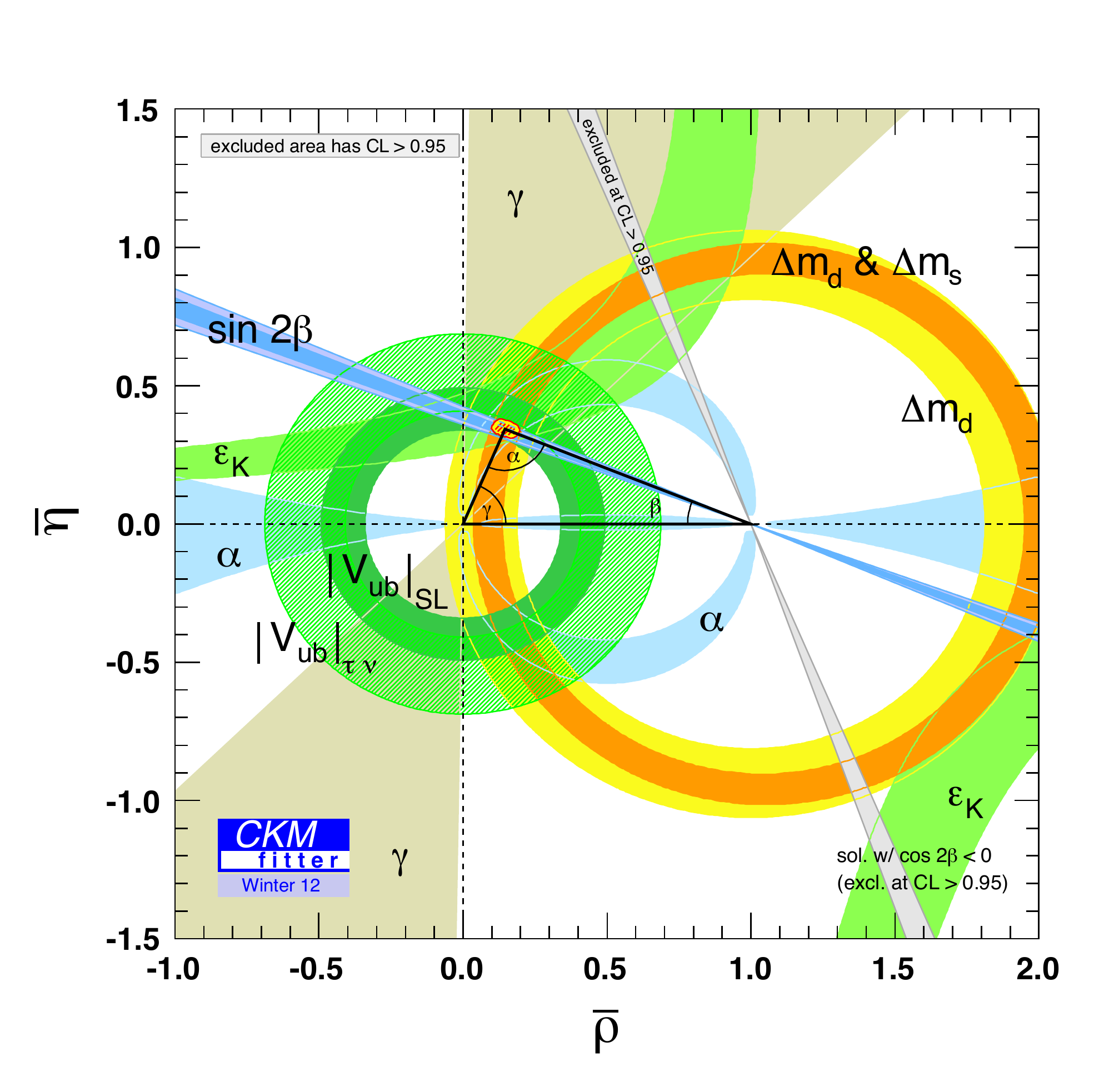}
\includegraphics[height=.33\textheight]{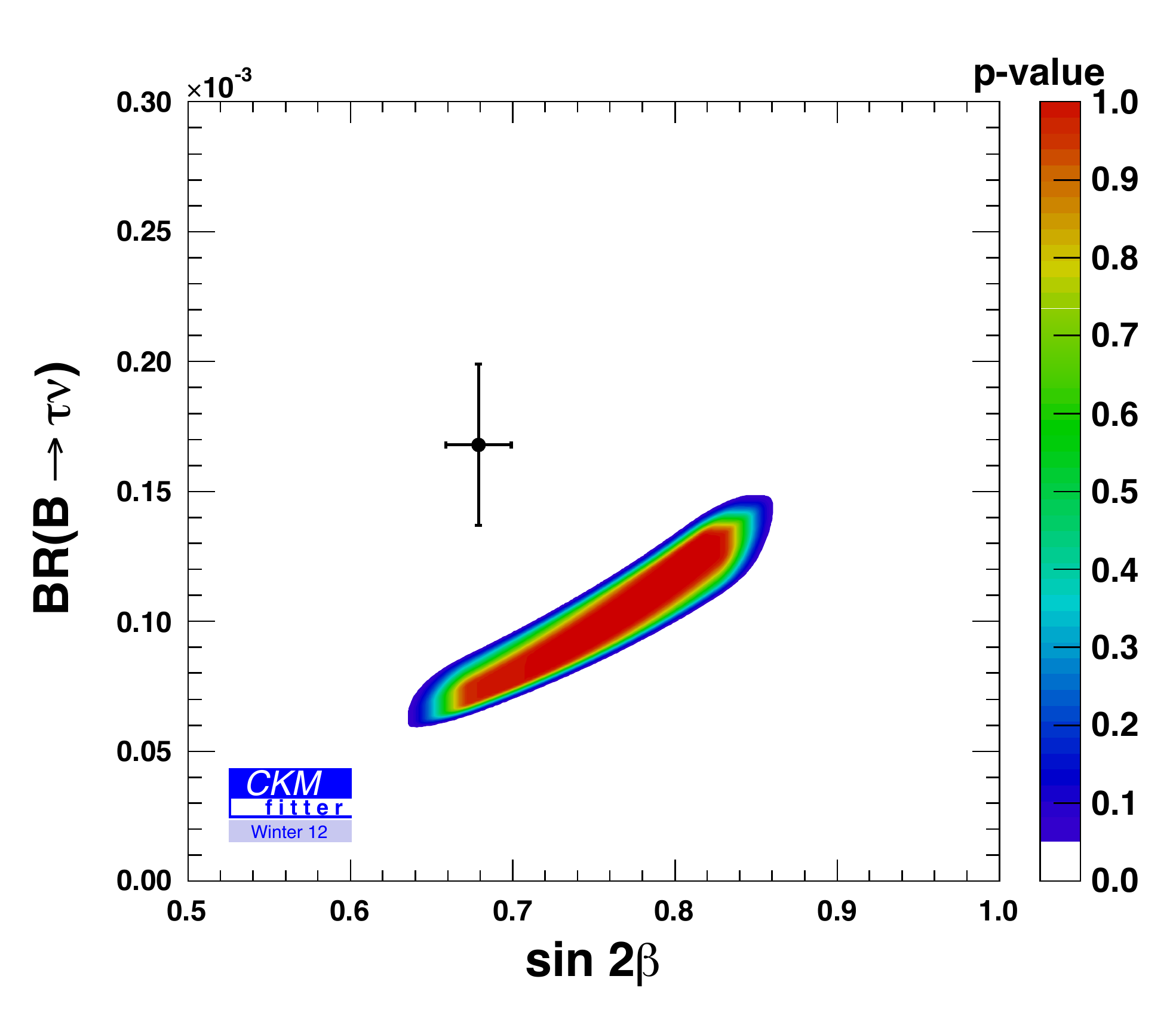}
\caption{On the left: 95\% CL constraints on the unitarity triangle from the SM global fit.\label{fig:UTriangle}. On the right: discrepancy between the measured values of $[\sin 2\beta, Br(B\to\tau\nu)]$ (cross) and the preferred values by the SM global fit (coloured region). \label{fig:sin2betavsBtaunu}}
\end{center}
\end{figure*}

The fit results  in the  $(\bar\rho,\bar\eta)$ plane are shown in Fig.~\ref{fig:UTriangle}, with the four real parameters describing the CKM matrix at 68\% CL:
\begin{eqnarray}
A&=&0.812^{+0.015}_{-0.022}\,, \quad
\lambda=0.2254^{+0.0006}_{-0.0010}\,, \\
\bar\rho &=&0.145^{+0.027}_{-0.027}\,,\quad
\bar\eta=0.343^{+0.015}_{-0.015}\,.
\end{eqnarray}
Even though there is a good overall agreement, the SM global fit exhibits a discrepancy between its various observables~\cite{Charles:2011va}. Indeed, the global fit $\chi^2_{min}$ drops by $2.8\sigma$ if $\sin 2\beta$ or $B\to\tau\nu$ is removed from the inputs. Before claiming any effect from New Physics, one should consider possible sources for such a discrepancy within the SM. On the experimental side, both $\sin 2\beta$ or $B\to\tau\nu$ showed a good 
agreement between Babar and Belle. On the theoretical side,
the issue is not restricted to the value of $f_{B_d}$ since one observes a $2.9\sigma$ discrepancy from the ratio
$BR(B\to\tau\nu)/\Delta m_d$, indicating that one needs to modify $f_{B_d}$ without affecting $f_{B_d}\sqrt{B_{B_d}}$.
Three solutions for this discrepancy could be: \emph{a)}
a change in the measurement of $B\to\tau\nu$~\footnote{This is in particular suggested by the recent measurement of $BR(B\to\tau\nu)$ from Belle presented at ICHEP 2012~\cite{Adachi:2012mm}. We do not include this result in these proceedings which correspond to the results available in Spring 2012. We invite the reader to 
browse through Ref.~\cite{ckmfitter} to learn more on the 
 impact of this new result.}, \emph{b)} a correlated change in lattice estimates of $f_{B_d}$ and $B_{B_d}$, with a value of $B_{B_d}$ much smaller than 1 (vacuum saturation approximation),
\emph{c)} the presence of NP affecting the interpretation of the observables in terms of the CKM parameters.

\section{$B_s\to \mu\mu$}

The dileptonic decay of the $B_s$ meson is among the most appealing
laboratories to study scalar couplings in addition to the SM ones, and complements the current effort in measuring 
$B\to K(^*)\mu\mu$ observables. The experimental limits set by the Tevatron and LHC
experiments on the dimuonic branching ratios~\cite{Aaltonen:2011fi,Aaij:2012ac,Chatrchyan:2012rg} are 
already giving significant constraints on scenarios beyond the Standard 
Model. A main limitation in the current prediction arises from the
knowledge of the decay constant $f_{B_s}$, as seen from
the master formula: 
\begin{eqnarray}
&&\!\!\!\!\!\!\!\!\!\!\!\!\!\!\!\!\!\!BR[ B_{s} \to \ell^+ \ell^-]_{\rm SM}  = \frac{G_F^2 \alpha_{em}^2  f_{B_{s}}^2 m_{\ell}^2  m_{B_{s}} \tau_{B_{s}} }{ 16 \pi^3 \sin^4\theta^{\rm eff}_W } \\
 &&\times \sqrt{ 1 - \frac{4m_{\ell}^2}{m_{B_{s}}^2} } |V^*_{tb}V_{t(s)}|^2\ Y^2\left(\frac{\bar{m}_t^2(\mu_t)}{m_W^2}\right),\qquad\qquad\nonumber
\label{eq_btoll}
\end{eqnarray}	
	 where $Y$ is the  Inami-Lim function at NLO in $O(\alpha_s)$~\cite{Buchalla:1995vs}.
	 $\bar{m}_t$ is the top quark mass defined in the $\overline{{\textrm{MS}}}$ scheme, considered at a scale $\mu_t=O(\bar{m}_t)$, $G_F$ is  the Fermi constant, $\sin^2\theta^{\rm eff}_W$ the electroweak mixing angle, and $\alpha_{em}$ the electromagnetic coupling constant at the $Z$ pole, where we use the tree-level formula~\cite{ALEPH:2005ab}:
\begin{equation}
\alpha_{em}(M_Z^2)=\frac{1}{\pi}\sqrt{2} G_F M_W^2 \sin^2\theta_W=1/127.46
\end{equation}
There are higher-order corrections to this formula, both in the electroweak and strong sectors, which we
account for by varying the renormalisation scale $\mu_t$ between $\bar{m}_t/2$ and $2\bar{m}_t$. The global SM fit yields an accurate prediction for the branching ratio:
\begin{equation}\label{eq:bsmumuSM}
BR[ B_s \rightarrow \mu^+ \mu^-]_{\rm SM}=(3.64^{+0.21}_{-0.32})\cdot 10^{-9}
\end{equation}
If we compare this prediction with other values in the literature~\cite{Buras:2010mh,Buras:2012ru,UTFit}, we find predictions spanning from 3.2 to 3.6, which may seem worrying considering that LHCb and CMS bounds ($4.5\cdot 10^{-9}$ and $7.7\cdot 10^{-9}$ at 95\% CL respectively) are getting close to these values. A first source of discrepancy, discussed in Ref.~\cite{Buras:2010mh}, corresponds to the choice of schemes for the electroweak mixing angle and the top quark mass. Any scheme can be chosen a priori, but one would like to pick up the one minimising the higher-order electroweak corrections. But the set of NLO electroweak corrections to this process is not wholly known, so that we do not know the preferred scheme, and different schemes seem to lead to a significant variation in the central values.
The case of $K\to\pi\nu\bar\nu$, where the NLO electroweak corrections are known, led the authors of Ref.~\cite{Buras:2010mh} to prefer the scheme used here. However this leaves open the question of the systematic uncertainty due to  our lack of knowledge about electroweak NLO corrections, which is not accounted for in the current predictions of the branching ratio [including Eq.~(\ref{eq:bsmumuSM})].

In any case, the difference between refs.~\cite{Buras:2010mh,Buras:2012ru} and our result cannot be ascribed to a difference of schemes. But there are also differences in the inputs used in these references and in Eq.~(\ref{eq:bsmumuSM}). Actually, when looking at the prediction from the SM global fit presented in Eq.~(\ref{eq:bsmumuSM}), one should take into account the fact that the global fit constrains not only the value of the CKM matrix elements, but also some of the hadronic inputs. Indeed, $f_{B_s}^2$ is constrained indirectly by $\Delta m_s$ and $B_{B_s}$ as they are both known with a better accuracy from our average  of lattice QCD results: the respective relative uncertainties for our inputs are 6~\%, 3 \% and 0.2 \% for $f_{B_s}^2$, $B_{B_s}$ and $\Delta m_s$. This leads the
global fit to predict $f_{B_s}=234^{+3}_{-8}$ MeV more accurately that the input in  Table~\ref{tab:inputs}.
The difference between the results of the SM global fit and Ref.~\cite{Buras:2010mh} for $BR(B_s\to\mu\mu)$ can be illustrated by taking the ratio $\rho=BR[B_s\to\mu^+\mu^-]_{SM}/\Delta m_s$ to eliminate $f_{B_s}$ in favour of $B_{B_s}$ and $\Delta m_s$:
\begin{equation}
\rho=\eta_Y^2\frac{6\pi}{\eta_B}\left(\frac{\alpha_{em}}{4\pi\sin^2\theta_W}\right)^2
\frac{m_\mu^2}{m_W^2}\frac{\tau_{B_s}}{\hat{B}_{B_s}} \frac{Y^2(x_t)}{S(x_t)}
\end{equation}
and insert our best-fit values for the inputs, to be compared with Ref.~\cite{Buras:2010mh}:
\begin{center}
\begin{tabular}{ccc}
Value         & Winter 2012 CKMfitter &  \cite{Buras:2010mh}\\
$\hat{B}_{B_s}$ & 1.252 & 1.33 \\
$\bar{m}_t(\bar{m}_t)$ (GeV) &  165.1 &163.5 \\
$\Delta m_s$ (ps$^{-1}$) & 17.73 & 17.77\\
$\tau_{B_s}$ (ps) & 1.472 & 1.425\\
$Br(B_s\to\mu\mu)$ & $3.6\cdot 10^{-9}$  & $3.2\cdot 10^{-9}$
\end{tabular}
\end{center}
Hence, the choice of inputs in Ref.~\cite{Buras:2010mh} and our improved knowledge of hadronic inputs due to the global fit yields the span of predictions in the literature.

Let us add that one should also include the effect of $\Delta\Gamma_s\neq 0$ to compare the theoretical branching ratio (computed at $t=0$, i.e., without mixing) and the measurement at hadronic machines (integrating over time and thus including mixing)~\cite{DescotesGenon:2011pb,DeBruyn:2012wj,DeBruyn:2012wk}. In the case of the Standard Model, the former should be 9\% below the latter, bringing the LHCb upper limit even closer to the SM value predicted from the global fit Eq.~(\ref{eq:bsmumuSM}).

\section{New Physics in $\Delta F=2$ operators: $M_{12}$}

In addition to the discrepancy between $B\to\tau\nu$ and $\sin 2\beta$ present
at the time of the Winter 2012 update, a large set of information has been obtained concerning
 $B_s\bar{B}_s$ mixing, as well as intriguing results for
 the dimuon charge asymmetry. It is interesting in this context to look for NP in mixing, i.e., $\Delta B=2$ operators, assuming that tree decays are not affected by NP effects~\cite{Lenz:2010gu,Lenz:2012az}.
$B_q\bar{B}_q$ oscillations (with $q=d$ or $q=s$) are described by a
Schr\"odinger equation with an evolution matrix for $|{B_q(t)}\rangle,|{\bar{B}_q (t)}\rangle$ of the form
$M^q - i \Gamma^q/2$,
with the hermitian mass and decay matrices $M^q$ and $\Gamma^q$. 

The physical eigenstates $|{B^q_H}\rangle$ and $|{B^q_L}\rangle$ with masses
$M^q_H,\,M^q_L$ and decay rates $\Gamma^q_H,\,\Gamma^q_L$ are obtained
by diagonalizing $M^q-i \, \Gamma^q/2$.  
The description of the  $B_q\bar{B}_q$ oscillations involves the three physical quantities $|M_{12}^q|$,
 $|\Gamma_{12}^q|$ and the CP phase $\phi_q = \arg(-M_{12}^q/\Gamma_{12}^q)$.
 The average $B_q$ mass and
width is $M_{B_q}$ and $\Gamma_{B_q}$, and the mass and width differences
between $B^q_L$ and $B^q_H$ can be written as
\begin{eqnarray}
\Delta M_q &=& M^q_H -M^q_L =  2\, |M_{12}^q|,\\
\Delta\Gamma_q &=& \Gamma^q_L-\Gamma^q_H  =
        2\, |\Gamma_{12}^q| \cos \phi_q, \label{dmdg}
\end{eqnarray}
A third quantity probing mixing is the semileptonic CP asymmetry,
\begin{equation}
a^q_{\rm fs} =
2\, \left( 1- \left| \frac{q}{p}\right| \right)=    {\rm Im}\  \frac{\Gamma_{12}^q}{M_{12}^q}
    = \frac{|\Gamma_{12}^q|}{|M_{12}^q|} \sin \phi_q
  \label{defafs}
\end{equation}
$a_{\rm fs}^q$ is the CP asymmetry in flavour-specific $B_q\to f$
decays ($f$ being chosen so that the decays $\bar B_q \to f$ and $B_q \to \bar f$ are forbidden), denoted $a_{\rm SL}^q$ for $f$ semileptonic.

These quantities are expected to be affected by NP differently.
While $M_{12}^q$ coming from box diagrams is very sensitive to NP 
both for $B_d$ and $B_s$, $\Gamma_{12}^q$ stems from 
tree-level decays and possible NP effects are generally expected to be smaller 
than the hadronic uncertainties from decay constants and bag parameters.
Assuming that NP does not enter tree-level 
decays (a more specific definition can be found in Ref.~\cite{Lenz:2010gu}), a first natural model-independent approach to NP in $\Delta F=2$ consists in changing the magnitude and/or the phase 
of $M_{12}^q$ only. It is convenient to define the NP complex parameters $\Delta_q$ and 
$\phi^\Delta_q$ ($q=d,s$) as $M_{12}^q \equiv  M_{12}^{{\rm SM},q} \cdot  \Delta_q$, with
\begin{equation}
 \Delta_q =  |\Delta_q| e^{i \phi^\Delta_q}\,,\qquad
\phi_q = \phi_q^{{\rm SM}} + \phi^\Delta_q . 
\end{equation}
New Physics in $M_{12}^q$ will not only affect the neutral-meson mixing parameters, but also the
time-dependent analyses of decays corresponding to an interference
between mixing and decay.  

The following inputs of the SM global fit are considered to be free from NP 
contributions in their extraction from experiment: $|V_{ud}|$, $|V_{us}|$,
$|V_{ub}|$, $|V_{cb}|$ and $\gamma$. We also assume that the leptonic decay $B\to \tau \nu$ is SM-like. Under these assumptions, a reference 
unitarity triangle can be constructed, with two solutions for the apex of the unitarity triangle (corresponding to the usual solution and the symmetric one with respect to the origin of the $(\bar\rho,\bar\eta)$ plane). On the other hand, we use several observables, affected by NP in mixing, to determine $\Delta_d$, $\Delta_s$:
the oscillation frequencies $\Delta m_d$, $\Delta m_s$,  the lifetime difference $\Delta\Gamma_d$, the time dependent asymmetries related to $\phi_{B_d},\phi_{B_s}$, the
asymmetries $a^d_{\rm SL}$, $a^s_{\rm SL}$, $A_{\rm SL}$, and finally 
$\alpha$ (from interference between decay and mixing).

 In summer 2010 the CDF and D\O\ analyses of $B_s \to
J/\psi \phi$ pointed towards a large negative value of $\phi^\Delta_s$,
while simultaneously being consistent with the SM due to large
errors. A large $\phi^\Delta_s<0$ could  
  accommodate D\O\ large negative value for the semileptonic CP
  asymmetry reading $A_{\rm SL} = 0.6 a_{\rm SL}^d + 0.4 a_{\rm SL}^s$~\cite{Abazov:2011yk}
  in terms of the individual semileptonic CP asymmetries in the $B_d$
  and $B_s$ systems.
Moreover, the discrepancy between $B(B\to \tau \nu)$
and the mixing-induced CP asymmetry in $B_d \to J/\psi K$ 
can be removed with
$\phi_d^\Delta<0$.  The allowed range for $\phi_d^\Delta$ implies a
  contribution to $A_{\rm SL}$ with the right (i.e.\ negative)
sign.  In our 2010 analysis~\cite{Lenz:2010gu} we have
performed a simultaneous fit to CKM parameters, $\Delta_s$ and $\Delta_d$ in three generic scenarios in
which NP is restricted to $\Delta F=2$ flavour-changing neutral currents. 
The most general scenario (called Scenario I)  treated  $\Delta_s$ and $\Delta_d$ independently, corresponding to
NP with arbitrary flavour structure. At that time, we found an excellent fit in this scenario,
with all discrepancies alleviated through
$\Delta_{d,s}\neq 1$.
The recent LHCb measurement~\cite{LHCb:2011aa} of  
  the CP phase $\phi_s^{\psi\phi}$ from $A_{\rm CP}^{\rm mix} (B_s \to
  J/\psi \phi)$ does not permit large deviations of $\phi_s^\Delta$ from
  zero anymore, as indicated also by CDF~\cite{CDF:2011af}. 
In Ref.~\cite{Lenz:2012az}, we have reconsidered our NP scenarios with these additional data.
 
 \begin{figure*}[t]
  \includegraphics[height=.35\textheight]{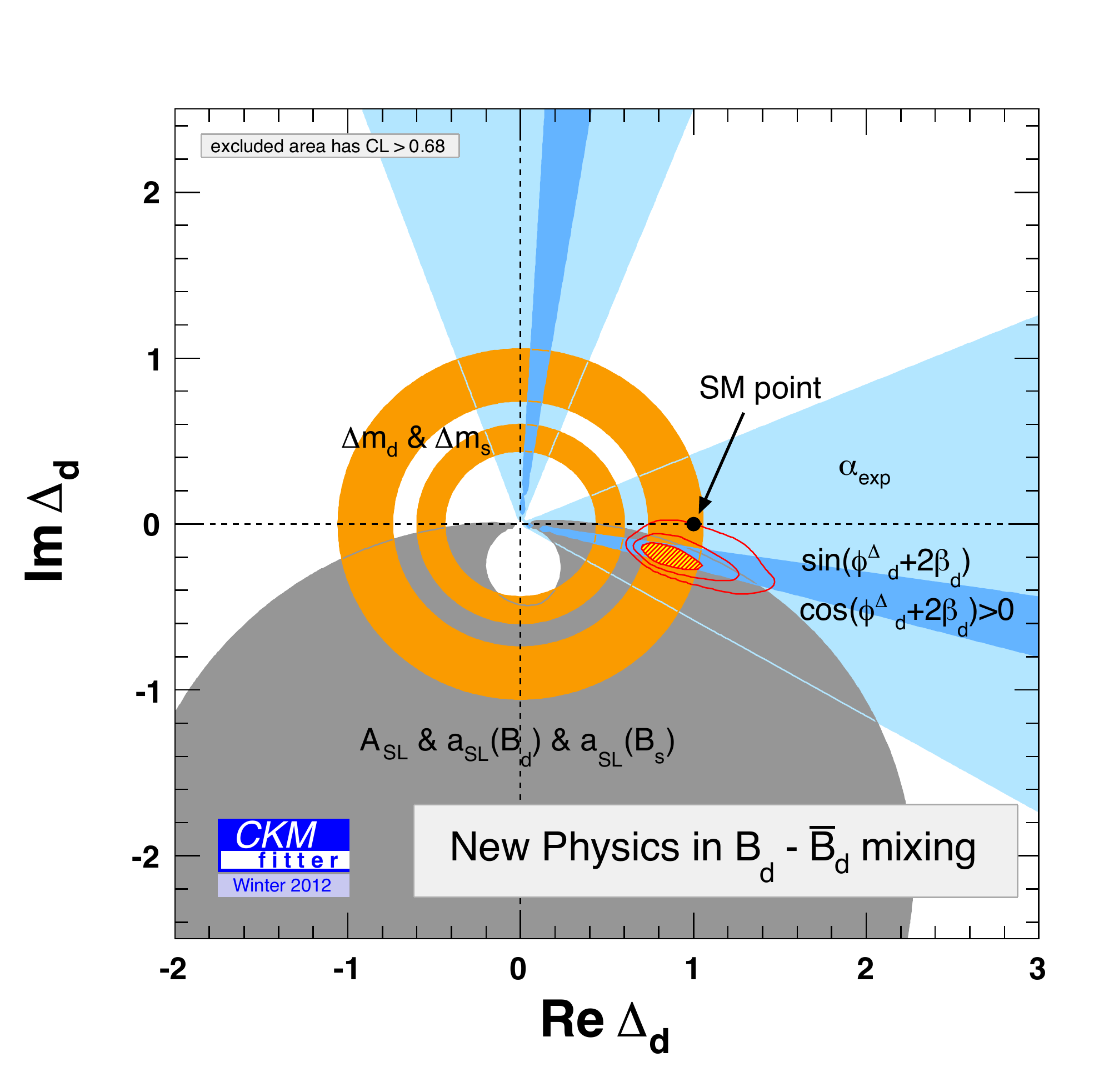}
 \includegraphics[height=.35\textheight]{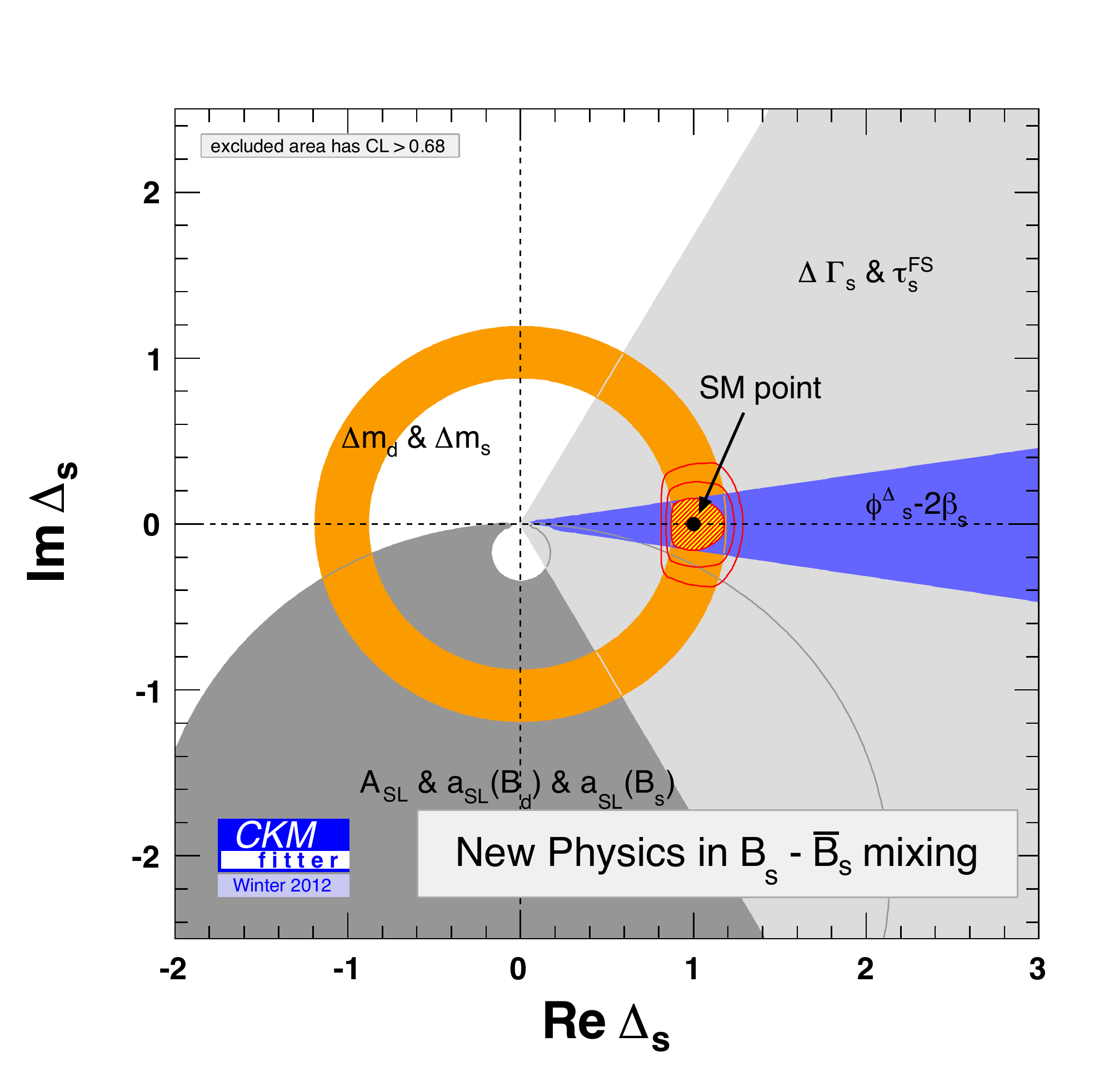}
 
  \caption{Constraint on the complex parameter $\Delta_d$ and $\Delta_s$.
                  For the individual constraints the coloured areas represent regions
                  with ${\rm CL} < 68.3~\%$. For the combined fit the red area shows
                  the region with ${\rm CL} < 68.3~\%$ while the two additional contour 
                  line inscribe the regions with ${\rm CL} < 95.45~\%$, 
                  and  ${\rm CL} < 99.73~\%$, respectively. We have
                  ${\rm Re}\ \Delta_d=0.823^{+0.143}_{-0.095}$, ${\rm Im}\  \Delta_d=-0.199^{+0.062}_{-0.048}$,
                  ${\rm Re}\ \Delta_s=0.965^{+0.133}_{-0.078}$, and ${\rm Im}\ \Delta_s=0.00\pm 0.10$.
                  The $p$-value for the 2D SM hypothesis 
                  $\Delta_d=1$ ($\Delta_s=1$) is 3.0~$\sigma$ (0.0 $\sigma$).
                  \label{fig-Deltad_scenario1}\label{fig-Delta_scenario1}
                  }
\end{figure*}

In Fig.~\ref{fig-Delta_scenario1}, we show the results in the
complex $\Delta_d$ and $\Delta_s$ planes.
$\Delta_d$ and $\Delta_s$ are
taken as independent in this scenario, but some of the
constraints correlate them, such that $A_{\rm SL}$ from the inclusive
dimuon asymmetry, and the ratio $\Delta m_d/\Delta m_s$. The
figures should be understood as two-dimensional projections of a
single multidimensional fit, and not as independent computations. On the left panel,
The constraint from $\Delta m_{d}$ in the
$\mbox{Re}{\Delta_d}-\mbox{Im}{\Delta_d}$ plane shows two
rings, related to the two allowed solutions for the reference triangle in the
$\bar\rho-\bar\eta$ plane. In the combined fit, the inner ring in the complex $\Delta_d$ plane is disfavoured, leading to an allowed region for $|\Delta_d|$  compatible with the
SM value $\Delta_d=1$ at the 3$\sigma$ level. The NP phase $\phi^\Delta_d$, mainly driven by the
discrepancy between $BR(B\to\tau\nu)$ and $\sin2\beta$, shows a clear deviation from the SM
 at the moment of the Winter 2012 update --  a fit without $BR(B\to\tau\nu)$ (or with an updated value following Ref.~\cite{Adachi:2012mm}) yields $\phi_{\Delta_d}$ in agreement with the SM.
On the right panel, we see that the recent LHCb results  have changed the
constraints on $|\Delta_s|$ and our fit  is significantly worse than in 2010~\cite{Lenz:2010gu}.
The value of $\phi_s^{\psi\phi}$ prevents large
contributions  to $A_{\rm SL}$ from  $B_s$ so that 
the two measurements do not show a good compatibility. We obtain
pull values for $A_{\rm SL}$ and $\phi_s^\Delta-2\beta_s$ of
 3.0$\,\sigma$ and 2.7$\,\sigma$ respectively (it was 1.2$\,\sigma$ and 0.5$\,\sigma$ in Ref.~\cite{Lenz:2010gu}).  

\begin{figure*}[t]
  \includegraphics[height=.33\textheight]{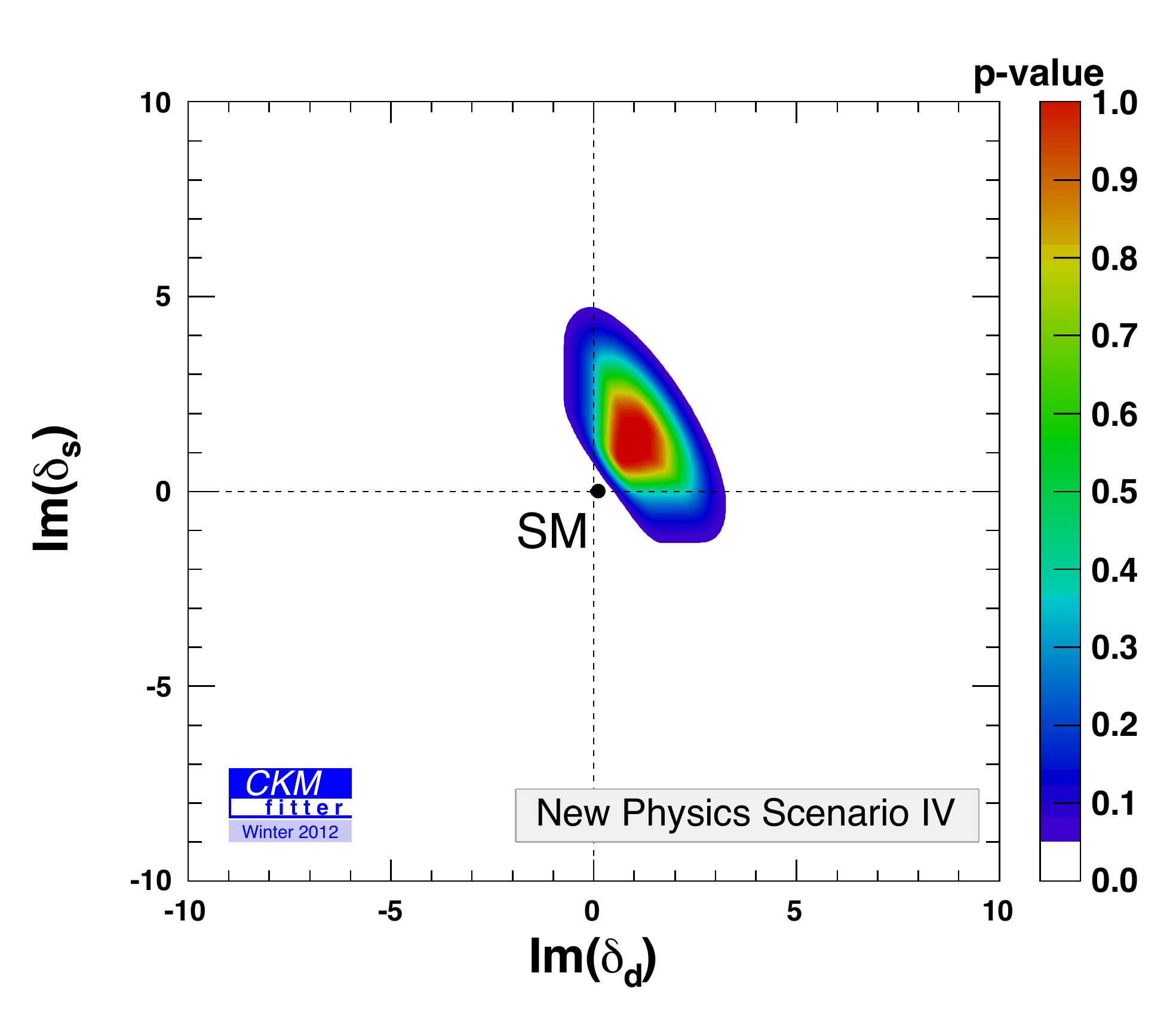}
 \includegraphics[height=.33\textheight]{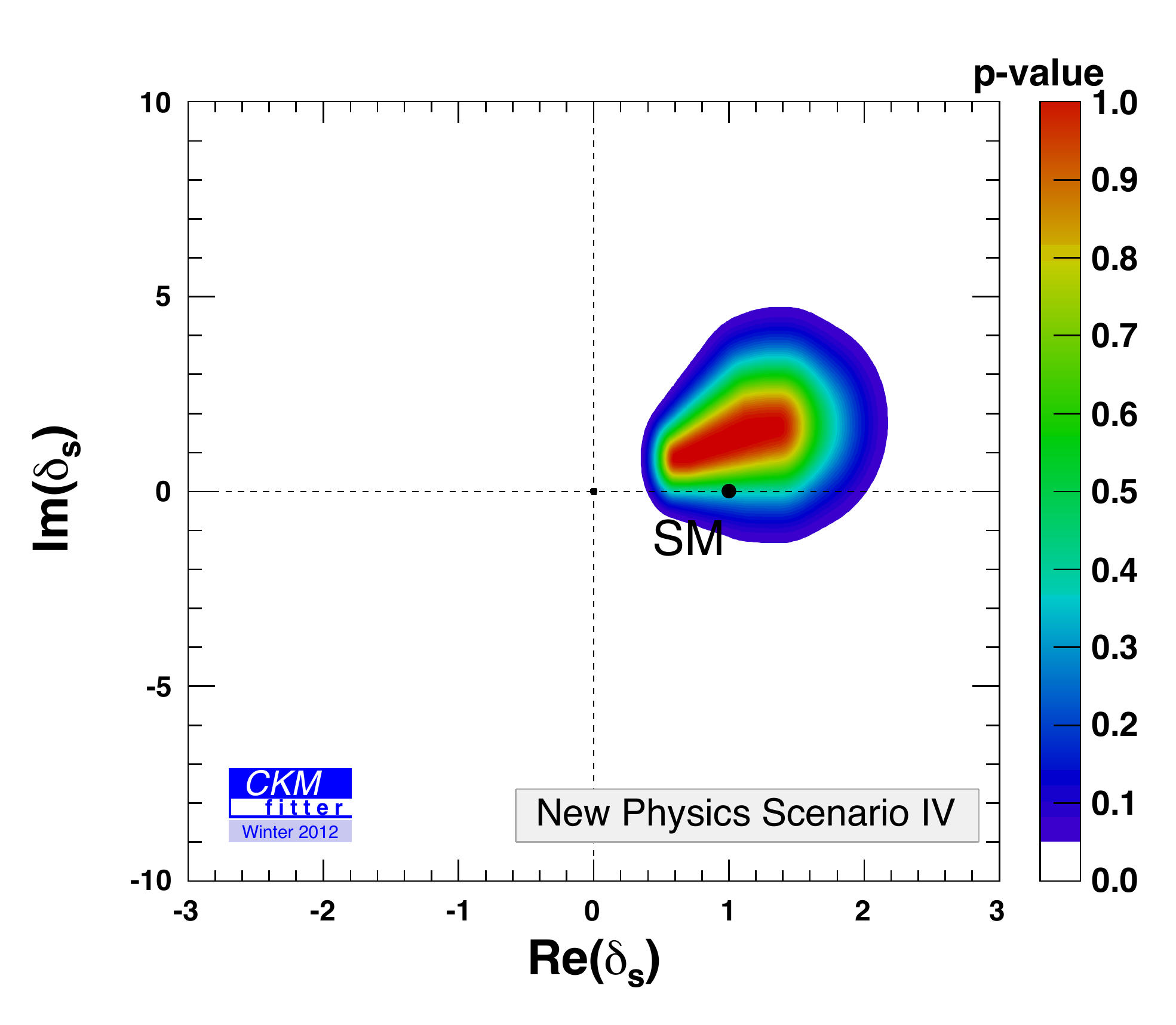}
\caption{Constraints on $({\rm Im}\ \delta_d,{\rm Im}\ \delta_s)$ and $({\rm Re}\ \delta_s,{\rm Im}\ \delta_s)$ in
    Scenario IV. The 68\% CL intervals are  
    ${\rm Im}\ \delta_d=0.92^{+1.13}_{-0.69},\ {\rm Im}\ \delta_s=1.2^{+1.6}_{-1.0}$. The best-fit
  values of the SM predictions are  $\delta_d^{\rm SM}=1 + 0.097\, i$ and $ \delta_s^{\rm SM}= 1 - 0.0057\, i$, and  the $p$-value for the 2D SM hypothesis ${\rm Im}\ \delta_{d,s}={\rm Im}\ \delta_{d,s}^{\rm SM}$  is
    3.2 $\sigma$.
}\label{fig-ScenarioIV}
\end{figure*}

\section{New Physics in $\Delta F=2$ operators: $\Gamma_{12}$}

In view of these results, new CP-violating contribution to $\Gamma_{12}^s$ can be advocated to explain the D\O\
measurement of $A_{SL}$~\cite{Abazov:2011yk}. It amounts to
postulating new $B_s$ decay channels with large branching fraction. This induces a SM deviation
in  $\Delta\Gamma_s$ as $\Gamma_{12}^s$ is dominated by the
CKM-favoured tree-level decay $b\to c\bar{c}s$. Any  new
decay mode increases the total $B_s$ width, 
whereas the measured
$\Gamma_s/\Gamma_d= 0.998 \pm 0.014 \pm 0.012 $ agrees
with the SM expectation $0\leq \Gamma_s/\Gamma_d-1\leq 4\cdot 10^{-4}$. The new interaction opens $b\to s$
decay modes affecting precisely measured inclusive $B_d$ and $B^+$ quantities \cite{Lenz:2010gu}.  New decays mediated
by a particle with mass $M>M_W$ will add a term of $O(M_W^4/M^4)$
to $\Gamma_{12}^s/\Gamma_{12}^{{\rm SM},s}$, whereas $\Delta_s$ normally
gets a larger contribution of $O(M_W^2/M^2)$. In models enhancing modes with
a fermion pair $(f,\bar f)$,  one can solve
this problem with chirality suppression. The extra contribution to
$M_{12}^s$ is down by a factor of $m_f^2/M^2$, while that to
$\Gamma_{12}^s$ is affected by the milder factor of $m_f^2/m_b^2$.
However quantities like $\Gamma_{d,s}$ will not be
chirality suppressed. It seems thus hard to add large NP effects to $\Gamma_{12}^s$ without spoiling the agreement of well-measured with the SM.

 Phenomenologically it is much easier to postulate
NP in $\Gamma_{12}^d$ as it stems from Cabibbo-suppressed decay modes like
$b\to c \bar{c}s d$. Chirality suppression can be invoked to
avoid problems with $M_{12}^d$, and there is no problem with inclusive decay observables like the
semileptonic branching fraction or the unmeasured $\Delta\Gamma_d$. In Ref.~\cite{Lenz:2012az}, we have studied a new scenario (Scenario IV) including the
possibility of NP in $\Gamma_{12}^{d,s}$.
  It enables NP in the $|\Delta F|=1$
  transitions contributing to $\Gamma_{12}^q$, but not in other $|\Delta
  F|=1$ quantities entering our fits, such as $BR(B\to \tau
  \nu)$. No new CP phase in $b\to c \bar{c} s$, which would
  change $\phi_{d,s}^\Delta$, is considered. 
We introduce new parameters 
\begin{equation}
\!\delta_q = \Gamma_{12}^q/M_{12}^q/{\rm Re}\ 
             (\Gamma_{12}^{{\rm SM},q}/M_{12}^{{\rm SM},q}) \qquad q={d,s}
             \end{equation}
${\rm Re}\ \delta_q$, ${\rm Im}\ \delta_q$ amount to
$(\Delta\Gamma_q/\Delta M_q)/(\Delta\Gamma_q^{\rm SM}/\Delta M_q^{\rm SM})$ and $-a_{\rm
  SL}^q/(\Delta\Gamma_q^{\rm SM}/\Delta M_q^{\rm SM})$, respectively.  ${\rm Re}\ \delta_d$ is experimentally
only weakly constrained and  the correlation between 
${\rm Re}\ \delta_s$, ${\rm Im}\  \delta_d $ and ${\rm Im}\ \delta_s$ is shown in Fig.~\ref{fig-ScenarioIV}. The p-value of the 8D SM hypothesis $\Delta_d=\Delta_s=1$,  $\delta_{d,s}=\delta_{d,s}^{\rm SM}$
is  2.6 $\sigma$. Too large values
for $|\delta_s-\delta_s^{\rm SM}|$ would be in conflict with other
observables as explained above.  One can also study this scenario without NP in the $B_s$ sector ($\Delta_s=1$ and $\delta_s=\delta_{s,\mathrm{SM}}$). 
It could accommodate the main anomalies by improving the fit by  $3.3\sigma$, but with large contributions to $\Gamma^d_{12}$: ${\rm Im}\ \delta_d=1.60^{+1.02}_{-0.76}$.

Considering the difficulties to accommodate all inputs (in particular $B_s\to J/\psi\phi$ and the dimuon charge asymmetry, independently of the discrepancy between $BR(B\to\tau \nu)$ and $\sin(2\beta)$~\cite{ckmfitter,Charles:2011va,Adachi:2012mm}),
 new measurements of $a^d_{\rm SL}$,
$a^s_{\rm SL}$ and/or $a^s_{\rm SL}-a^d_{\rm SL}$ are necessary
to determine whether scenarios with NP in $\Gamma_{12}^d$
  and/or $\Gamma_{12}^s$ are a viable explanation for the discrepancies
  in neutral-meson mixing observables with respect to the SM, or if other avenues should be investigated (e.g., Ref.~\cite{DescotesGenon:2012kr}).

\nocite{*}
\bibliographystyle{elsarticle-num}

\end{document}